\documentclass[12pt]{article}
\pdfoutput=1
\usepackage{amsmath, amsfonts,euscript,bbm,mathrsfs,amssymb}
\usepackage{epsfig, cite}
\usepackage{color}
\usepackage{ifthen}
\usepackage{graphicx}
\usepackage{setspace}

\addtocontents{toc}{\protect\setcounter{tocdepth}{1}}

\newcommand{\ud}{\mathrm{d}}
\newcommand{\vd}{\text{d}}
\begin{document}

\bigskip
\bigskip\bigskip\bigskip\bigskip\bigskip
\bigskip\bigskip\bigskip\bigskip\bigskip

\centerline{\Large Partial Transmutation of Singularities in Optical Instruments}
\bigskip
\bigskip
\bigskip
\centerline{\bf Janos Perczel$^{1}$ and Ulf Leonhardt$^2$}
\medskip
\centerline{School of Physics and Astronomy}
\centerline{University of St Andrews}
\centerline{North Haugh}
\centerline{St Andrews, KY16 9SS, UK}
\medskip
\centerline{\it $^1$jp394@st-andrews.ac.uk}
\centerline{\it $^2$ulf@st-andrews.ac.uk}

\bigskip
\bigskip
\bigskip
\bigskip
\bigskip
\bigskip
\begin{abstract}
 
Some interesting optical instruments such as the Eaton lens and the Invisible Sphere require singularities of the refractive index for their implementation. We show how to transmute those singularities into harmless topological defects in anisotropic media without the need for anomalous material properties.

\end{abstract}
\newpage       

\section{Introduction}

Recent developments in the field of transformation optics \cite{Service:2010,Dolin:1961,Greenleaf:2003,Leonhardt:2006a,Pendry:2006,Leonhardt:2006b,Shalaev:2008,Leonhardt:2009c,Leonhardt:2010} and metamaterials \cite{Milton:2002,Smith:2004,Costas:2007,Sarychev:2007,Cai:2009,Capolino:2009} have led to the proposal of a number of remarkable optical instruments, including the Eaton Lens \cite{Hannay:1993,Eaton:1952,Kerker:1969,Ma:2009}, the Invisible Sphere \cite{Hendi:2006,Minano:2006,Leonhardt:2010} and invisibility cloaks \cite{Leonhardt:2006a,Pendry:2006,Leonhardt:2006c,Leonhardt:2009,Tycetal:2010}.
However, the implementation of these devices often requires physical parameters that are very difficult to achieve in practice. For example, a requirement for some is that the speed of light goes to zero (e.g. at the geometric centre of the Invisible Sphere and the Eaton Lens) or to infinity (e.g. in conformal invisibility cloaking \cite{Leonhardt:2006a,Leonhardt:2006c}). Such extremal requirements can be usually met only for discrete resonant frequencies, which severely restricts the applicability of these devices \cite{Tycetal:2010}.\\
A method has been proposed by Tyc and Leonhardt to transmute the refractive index profiles of singular optical devices into equivalent non-singular profiles with a slight anisotropy \cite{Tycetal:2008}. This technique is based on the expansion of space around a singularity with a non-constant factor to bring all index values within a finite range. The transmutation of singularities was a crucial step towards the practical realisation of singular optical devices (e.g. the Eaton lens \cite{Ma:2009}). However, the method of transmutation established in \cite{Tycetal:2008} always leads to anomalous eigenvalues for the permittivity and permeability tensors (i.e. values that are less than unity), which means that the transmuted profiles can be implemented only for discrete resonant frequencies \cite{Tycetal:2010}. In order to avoid anomalous transmuted index values a modification of the original method was suggested by Danner \cite{Danner:2010}. The modified technique is based on changing the expansion factor derived in \cite{Tycetal:2008} to make all transmuted values greater than unity.\\
In this paper we show that a completely non-anomalous transmutation can also be achieved by making the transmutation derived in \cite{Tycetal:2008} partial. This modification makes transmutation a local operation, which can be advantageous in a number of cases. First, partial transmutation minimises the anisotropic region in transmuted singular devices. Second, this technique can be applied to devices with potentially any number of isolated divergent singularities. Third, partial transmutation leaves the major part of the refractive index profile intact, which allows us to exploit desirable features of the non-transmuted isotropic refractive index profile that would otherwise disappear in a full transmutation done according to \cite{Tycetal:2008} or \cite{Danner:2010}. For our purposes the third point is the crucial one, which enables us to produce a partially transmuted Invisible Sphere, which can  be used as an invisible background profile for invisibility cloaks to avoid superluminal propagation while achieving electromagnetic cloaking. For a full account on the application of the partially transmuted Invisible Sphere as a background for invisibility cloaks see \cite{Leonhardt:2011a}.

\section{Examples}

First, we introduce the idea behind partial transmutation through the examples of the Invisible Sphere and the Eaton Lens.
The Invisible Sphere is an optical lens that is invisible in the limit of geometrical optics \cite{Hendi:2006,Minano:2006,Leonhardt:2010}. Each incident ray completes a loop in the sphere and emerges in the direction of its entry (Fig.~1A), creating the illusion that space is empty. This device has a spherically symmetric refractive index profile given by \cite{Minano:2006}

\begin{eqnarray}
n(r)=\bigg( Q-~1/(3Q)\bigg)^2 \:\,  \text{if $r \leqslant 1$, }   \text{and} \;\,
n(r)=1 \;\, \text{ if $r > 1$,}
\end{eqnarray}

where

\begin{equation}
Q(r)=\sqrt[3]{-\frac{1}{r}+\sqrt[2]{\frac{1}{r^2}+\frac{1}{27}}}.
\end{equation}

In the expression above we assume that the device has unit radius. It is easy to see that $n(r)$ tends to infinity as $r \to 0$ and that $n(r)$ diverges as $\sim$$r^{-2/3}$ near the geometric centre of the sphere \cite{Leonhardt:2010}. 
The diverging refractive index profile implies that the speed of light goes to zero at the geometric centre, which is extremely difficult to achieve in practice. However, we could follow the method described in \cite{Tycetal:2008} to transmute the refractive index profile of the Invisible Sphere into an equivalent anisotropic profile by expanding the space around the singularity with a factor that increases continuously as we get closer to the singular point. Following the transmutation, all eigenvalues of the permittivity and permeability tensors will be finite and thus no singularity will be present. Fig.~1B shows the light trajectories in the transmuted profile.\\
However, calculations show that the complete transmutation of the profile of the device leads to anomalous eigenvalues for the permittivity and permeability tensors, i.e. to eigenvalues that are less than unity. This can be a difficulty since such a profile would work only for incident waves with frequencies that are close to the resonant frequencies of the implementing material \cite{Milonni:2005}. In order to avoid this difficulty, we perform a partial transmutation on the original profile of the Invisible Sphere. This means that we leave the outer region of the profile unaltered and expand space only in the immediate vicinity of the singularity within a given radius $b$. Fig.~1C shows the partially transmuted profile. Calculations show that if we start the transmutation of the inner region at a radius less than or equal to $r=0.2489$, all eigenvalues of the permittivity and permeability tensors will be greater than unity, yet finite.\\
\begin{figure}[h!]
\begin{center}
\includegraphics[width=2in]{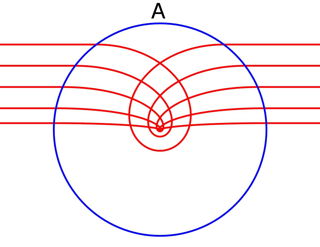}
\includegraphics[width=2in]{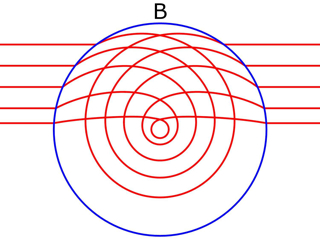}
\includegraphics[width=2in]{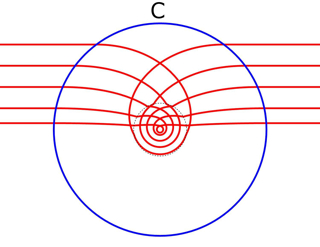}
\caption{A: Light trajectories in the Invisible Sphere. B: Trajectories in the completely transmuted Invisible Sphere. C: Trajectories in the partially transmuted Invisible Sphere, where the transmutation starts from r=0.2489.}
\end{center}
\end{figure}The partially transmuted Invisible Sphere can be used as an invisible background profile for anomalous invisibility cloaks \cite{Leonhardt:2011a}, because the cloak can be placed outside the transmuted region, where the refractive index is high. Thus the Invisible Sphere can raise the anomalous index values of the cloak above one, thereby making the implementation of the cloak practicable over a broad range of frequencies \cite{Leonhardt:2011a}.\\
Another important example of singular optical devices is the Eaton lens \cite{Hannaya:1993,Eaton:1952,Kerker:1969,Ma:2009}, which is an omnidirectional retroreflector (Fig.~2A). It also has a spherically symmetric refractive index distribution given by

\begin{eqnarray}
n(r)=\sqrt{2/r-1}\;\,  \text{if $r \leqslant 1$, }   \text{and} \;\,
n(r)=1 \;\, \text{ if $r > 1$}. 
\end{eqnarray}

In the expression above we assume that the device has unit radius. The refractive index goes to infinity as $r \to 0$. It can also be shown that $n(r)$ diverges as $\sim$$r^{-1/2}$ near the geometric centre of the the lens \cite{Leonhardt:2010}. We can follow the same procedure for the Eaton Lens as for the Invisible Sphere to obtain the completely transmuted version of the lens which has finite but anomalous refractive index values (Fig.~2B). By performing partial transmutation at a radius less than or equal to $r=0.3585$, we can arrive to an anisotropic profile whose tensor eigenvalues are greater than unity and finite at all points (Fig.~2C).
\begin{figure}[h!]
\begin{center}
\includegraphics[width=2in]{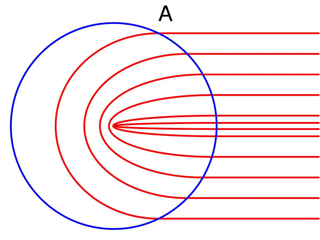}
\includegraphics[width=2in]{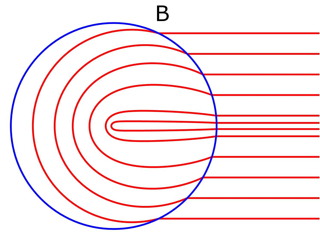}
\includegraphics[width=2in]{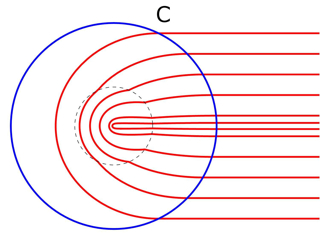}
\caption{A: Light trajectories in the Eaton Lens. B: Trajectories in the completely transmuted Eaton Lens. C: Trajectories in  the partially transmuted  Eaton Lens, where transmutation starts from r=0.3585.}
\end{center}
\end{figure}
In fact, it turns out to be generally true that for a profile with an isolated singularity and with asyptotic form $n(r)\sim$$r^{p}$ (where $-1<p<0$) an equivalent partially transmuted anisotropic refractive index distribution can be found, where all values of the permittivity and permeability tensors are greater than unity and finite, and whose overall influence on the light trajectories is equivalent to that of the non-transmuted profile.

\section{Method}

In the previous section we argued that refractive index profiles with isolated divergent singularities can be changed into equivalent non-anomalous finite profiles by partial transmutation. In this section we establish this result in a more general setting.
Let us consider profiles that have singularities that behave like $\sim$$r^{p}$ with $-1<p<0$ in the asymptotic limit. This implies that the refractive index goes to infinity and the speed of light goes to zero as $r \to 0$. It was shown in \cite{Tycetal:2008} that we can give enough space for the propagation of light and bring all refractive index values into a finite range if we transform the radial coordinate according to $R=r^{p+1}$ and leave the angular directions unchanged. The expansion factor between the virtual and physical space than becomes $\eta_{R}=\vd R/ \vd r=(p+1)r^{p}$, which implies that light will move $\eta_{R}$ times faster in physical space than in virtual space. Therefore, the refractive index will be $n_{r}=1/(p+1)$ in the radial direction and $n_{\theta}=n_{\phi}=1$ in the angular directions in physical space. Consequently, the singularity is effectively removed.\\
However, following the transmutation another problem emerges. Since $n^2=(\text{det}\,\varepsilon)\varepsilon^{-1}$ (where $n$ and $\varepsilon$ are the refractive index and permittivity tensors respectively) \cite{Leonhardt:2010}, it follows that $\varepsilon=(\text{det}\,n)n^{-2}$. Therefore, $\varepsilon_{r}=(p+1)$ and $\varepsilon_{\phi}=\varepsilon_{\theta}=(p+1)^{-1}$. Consequently, for $-1<p<0$  the radial component of the permittivity tensor of the profile will always be less than one, giving rise to anomalous dispersion \cite{Milonni:2005}, which severely restricts the applicability of the device.\\ 
In order to avoid such anomalous refractive indices, instead of transmuting the entire profile of the singular optical instrument, we leave the outer region of the profile intact and transmute only the immediate vicinity of the singularity within a given radius $b$ (Fig.~2C). In this case, the expansion of the radial coordinate in the transmuted region is still according to the power law: $R\sim$$r^{p+1}$, but now we will need to include a normalisation factor in front $R=\frac{1}{\mathcal{N}}\cdot$$r^{p+1}$ to make sure that $R(b)=b$ (i.e. that the transformation does not introduce impossible jumps in the trajectories), where $b$ is the outer boundary of the transmutation. Consequently, the expansion factor will now be $\eta_{R}=\vd R/\vd r=\mathcal{N}^{-1}(p+1)$$r^{p}$, giving rise to refractive index values $n_{r}=\mathcal{N}/(p+1)$ and $n_{\theta}=n_{\phi}=\mathcal{N}$. It is easy to show that $\mathcal{N}(b)=b^{p}$, which tends to infinity as $b \to 0$ (Fig.~3). Since $n_{r}$, $n_{\theta}$, and $n_{\phi}$ are all proportional to $\mathcal{N}(b)$, it follows that their value can be increased to any arbitrary finite value by choosing a sufficiently small b for the radius of the transmutation and thus the area of the transmutation can be reduced to an arbitrary small area.

\section{Conclusion}

In conclusion, we have established that for optical instruments that have isolated singularities diverging as $\sim$$r^{p}$ (where $-1<p<0$) an equivalent partially transmuted anisotropic refractive index can be found whose tensor eigenvalues are always greater than unity and finite, and whose overall influence on the light trajectories is completely equivalent to that of the non-transmuted profile. This method can be applied to optical devices with any number of singularities.

\section*{Acknowledgements}

We thank Natalia Korolkova, William Simpson, and Tomas Tyc for their valuable comments and help. Our work was supported by the Engineering and Physical Sciences Research Council and the Royal Society.

\appendix

\section{Appendix}

\subsection{Material properties}

We calculate the material properties of the transmuted optical devices in physical space by the method described in \cite{Leonhardt:2006b,Leonhardt:2009c,Leonhardt:2010}. 
We consider an optical device with a refractive index distribution that is spherically symmetric with $n(r)=r^{p}$ for $r \leqslant 1$ and $n(r)=1$ for $r>1$. We transform the radial coordinate by the following function

\begin{equation}
R(r)=\frac{1}{\mathcal{N}(b)} \, r^{p+1} \;\,  \text{if $r \leqslant b$, }   \text{and} \;\,
R(r)=r \;\, \text{ if $r > b$}.
\end{equation}

We leave the angular coordinates unchanged.
This represents a transformation from virtual space [described by the metric tensor $g_{i'j'}=\text{diag}(1,r^{2},r^{2} \text{sin}^{2}\theta)$] to physical space (described by $\gamma_{j k}=\text{diag}(1,R^{2},R^{2} \text{sin}^{2}\theta)$). The transformation matrix between the two spaces is is given by the Jacobian: $\Lambda^{i'}_{\,\, i}=\partial(r,\theta,\phi)/\partial(R,\theta,\phi)=\text{diag}(\vd r/\vd R,1,1).$
The mixed components of the permittivity and permeability tensors $\varepsilon^{i}_{\,\,k}$, $\mu^{i}_{\,\,k}$ can be calculated in physical space by the recipe given by equation (9) in \cite{Leonhardt:2006b} and  equation (32.1) in \cite{Leonhardt:2010}

\begin{equation}
\varepsilon^{i}_{\,\,k}=\mu^{i}_{\,\,k}=n\,\frac{\sqrt{g}}{\sqrt{\gamma}}g^{ij}\gamma_{j k} \text{,  \;\; } g_{ij}=g_{i'j'}\Lambda^{i'}_{\,\, i}\Lambda^{j'}_{\,\, j}.
\end{equation} 

In the expression above $g^{ij}$ is the inverse of $g_{ij}$, $g$ is the determinant of $g_{ij}$, $\gamma$ is the determinant of $\gamma_{ij}$ and $n$ is the refractive index. In the above equation we assume that the dielectric medium is impedance-matched, i.e.$\,\varepsilon^{i}_{\,\,k}=\mu^{i}_{\,\,k}$.
Substituting the relevant expressions into $(6.5)$ gives

\begin{equation}
\varepsilon^{i}_{\,\,k}=\mu^{i}_{\,\,k}=n\,\text{diag}\Big( \frac{r^2}{R^2}\frac{\ud R}{\ud r},\frac{\ud r}{\ud R},\frac{\ud r}{\ud R} \Big).
\end{equation} 

Substituting $n=r^{p}$ and $R=\frac{1}{\mathcal{N}(b)} \, r^{p+1}$ into the above expression gives the explicit expression for the mixed tensors in physical space:

\begin{equation}
\varepsilon^{i}_{\,\,k}=\text{diag}\Big(\varepsilon_{R},\varepsilon_{\phi},\varepsilon_{\theta}\Big)=\mu^{i}_{\,\,k}=\text{diag}\Big( \mathcal{N}(b)\, (p+1),\frac{\mathcal{N}(b)}{p+1},\frac{\mathcal{N}(b)}{p+1} \Big),
\end{equation} 

This result shows that the singularity has been effectively removed through the transmutation and that the tensor components are directly proportional to $\mathcal{N}(b)$. Therefore, by choosing a sufficiently small transmutation radius $b$, all tensor components can be brought above unity. If we start the transmutation from the boundary of the device $(b=1)$, then $\mathcal{N}=1$ and we recover the results derived in \cite{Tycetal:2008}.

\subsection{Invisible Sphere and Eaton lens}

Section 5. of \cite{Tycetal:2008} shows that transmutation is possible even when $n(r)$ has a general form that diverges like $r^{p}$ as $r \to 0$. For example, such a general refractive index profile can be transmuted by the function $R(r)=\frac{1}{n_{0}(b)}\int_{0}^{r}n(r')\ud r'$, where $n_{0}(b)$ is the normalisation constant.  This transmutation will keep the angular eigenvalues of $\varepsilon^{i}_{\,\,k}$ and $\mu^{i}_{\,\,k}$ constant and the radial eigenvalue will decrease slightly with increasing radius.
For the Invisible Sphere the integral gives the expression 

\begin{equation}
R(r)=\frac{1}{\mathcal{N_{IS}}(b)} \,\, f(r),\quad  \text{where} \quad f(r)=\dfrac{\int_{0}^{r}n(r')\ud r'}{\int_{0}^{1}n(r')\ud r'}\,.
\end{equation}

In the expression above the normalisation constant $\mathcal{N_{IS}}(b)$ was chosen such that $R(b)=b$ and $\mathcal{N_{IS}}(1)=1$. Therefore,  $\mathcal{N_{IS}}(b)=f(b)/b$. In Fig.~3 we plotted $\mathcal{N_{IS}}(b)$ in red as a function of $b$.\\
Since $\mathcal{N_{IS}}(b)$ increases as $b$ decreases and $n_{r}$, $n_{\theta}$, and $n_{\phi}$ are all proportional to $\mathcal{N_{IS}}(b)$, it is easy to see that by a sufficiently small choice of $b$ all tensor components can be raised above unity.
Calculations show that $b=0.2489$ is the largest and therefore optimal transmutation radius for which all the tensor eigenvalues of the Invisible Sphere are greater than or equal to one.  In Fig.~~4 we plotted the radial and angular eigenvalues for the partial transmutation with $b=0.2489$.\\
\begin{figure}[h!]
\begin{center}
\includegraphics[width=3in]{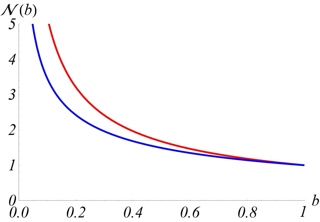}
\caption{The plot shows the normalisation constant for the Invisible Sphere (red) and the Eaton Lens (blue). Since $\mathcal{N}$ increases as $b$ decreases, by choosing a sufficiently small transmutation radius all entries of $\varepsilon^{i}_{\:j}$ and $\mu^{i}_{\:k}$ can be brought above one.}
\end{center}
\end{figure}
\begin{figure}[h!]
\begin{center}
\includegraphics[width=3in]{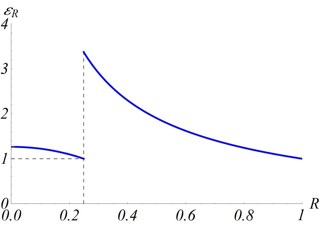}
\includegraphics[width=3.in]{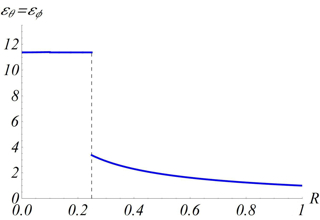}
\caption{The plots show the values of the radial and angular elements of the mixed tensors $\varepsilon^{i}_{\:j}=\mu^{i}_{\:k}$ of the partially transmuted Invisible Sphere, where $b=0.2489$. $\varepsilon_{R}$ changes between $1$ and $3.38$, and $\varepsilon_{\theta}$  and $\varepsilon_{\phi}$ change between $1$ and $11.38$. The refractive index profile remains unaffected outside the radius of the transmutation.}
\end{center}
\end{figure}
\begin{figure}[h!]
\begin{center}
\includegraphics[width=3in]{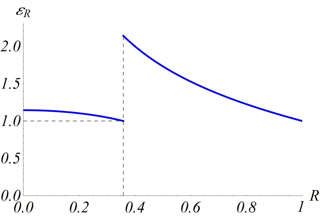}
\includegraphics[width=3in]{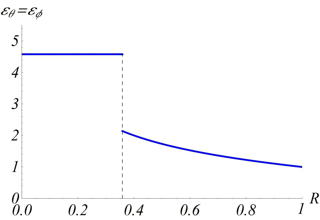}
\caption{The plots show the values of the radial and angular elements of the mixed tensor $\varepsilon^{i}_{\:j}=\mu^{i}_{\:k}$ of the partially transmuted Eaton Lens, where b=0.3585. $\varepsilon_{R}$ changes between $1$ and $2.14$, and $\varepsilon_{\theta}$  and $\varepsilon_{\phi}$ change between $1$ and $4.58$.}
\end{center}
\end{figure}

\newpage
For the Eaton Lens the integral gives \cite{Tycetal:2008}
\begin{equation}
R(r)=\frac{1}{\mathcal{N_{E}}(b)}\,f(r), \quad \text{where} \quad f(r)=\frac{4}{\pi+2} \, \bigg[\text{arcsin}\sqrt{\frac{r}{2}}+\sqrt{\frac{r}{2}\Big(1-\frac{r}{2}\Big)} \: \bigg].
\end{equation}
In the expression above $\mathcal{N_{E}}(b)$ ensures that $R(b)=b$ and $\mathcal{N_{E}}(1)=1$. We plotted $\mathcal{N_{E}}(b)$ in Fig.~3 in blue. For the Eaton Lens the optimal transmutation radius is $b=0.3585$.  Fig.~5 shows the radial and angular eigenvalues for the partial transmutation with $b=0.3585$.



\begin{thebibliography}{99}



\bibitem{Service:2010}
R. F. Service and A. Cho, Science {\bf 330}, 1622 (2010).

\bibitem{Dolin:1961}
L. S. Dolin, Isv. Vusov {\bf 4}, 964 (1961).

\bibitem{Greenleaf:2003}
A. Greenleaf, M. Lassas and G. Uhlmann, Math. Res. Lett. {\bf 10}, 685 (2003).

\bibitem{Leonhardt:2006a}
U. Leonhardt, Science {\bf 312}, 1777 (2006).

\bibitem{Pendry:2006}
J. B. Pendry, D. Schurig and D. R. Smith, Science {\bf 312}, 1780 (2006).

\bibitem{Leonhardt:2006b}
U. Leonhardt and T. G. Philbin, New J. Phys. {\bf 8}, 247 (2006).

\bibitem{Shalaev:2008}
V. M. Shalaev, Science {\bf 322}, 384 (2008).

\bibitem{Leonhardt:2009c}
U. Leonhardt and T. G. Philbin, Prog. Opt. {\bf 53}, 69 (2009).

\bibitem{Leonhardt:2010}
U. Leonhardt and T. G. Philbin, 
{\it Geometry and Light: The Science of Invisibility} (Dover, Mineola, 2010).



\bibitem{Milton:2002}
G. W. Milton, 
{\it The Theory of Composites}
(Cambridge University Press, Cambridge, 2002).

\bibitem{Smith:2004}
D. R. Smith, J. B. Pendry and M. C. K. Wiltshire, Science {\bf 305}, 788 (2004).

\bibitem{Costas:2007}
C. M. Soukoulis, S. Linden and M. Wegener, Science {\bf 315}, 47 (2007).

\bibitem{Sarychev:2007}
A. K. Sarychev, V. M. Shalaev, 
{\it Electrodynamics of Metamaterials}
(World Scientific, Singapore, 2007).

\bibitem{Cai:2009}
W. Cai and V. M. Shalaev, 
{\it Optical Metamaterials: Fundamentals and Applications}
(Springer, Berlin, 2009).

\bibitem{Capolino:2009}
{\it Theory and Phenomena of Metamaterials (Metamaterials Handbook)}
F. Capolino (ed.), (CRC Press, Boca Raton, 2009).


\bibitem{Eaton:1952}
J. E. Eaton, Trans. IRE Antennas Propag. {\bf 4}, 66 (1952).

\bibitem{Kerker:1969}
M. Kerker, 
{\it The scattering of light, and other electromagnetic radiation}
(Academic, New York, 1969).

\bibitem{Hannay:1993}
J. H. Hannay and T. M. Haeusser, J. Mod. Optics {\bf 40}, 1437 (1993).

\bibitem{Ma:2009}
Y. G. Ma, C. K. Ong, T. Tyc and U. Leonhardt, Nature Materials {\bf 8}, 639 (2009).



\bibitem{Minano:2006}
J. C. Mi\~nano, Opt. Express {\bf 14}, 9627 (2006).

\bibitem{Hendi:2006}
A. Hendi, J. Henn, and U. Leonhardt, Phys. Rev. Lett. {\bf 97}, 073902 (2006).


\bibitem{Leonhardt:2006c}
U. Leonhardt, New. J. Phys. {\bf 8}, 118 (2006).

\bibitem{Leonhardt:2009}
U. Leonhardt and T. Tyc, Science {\bf 323}, 110-112 (2009).

\bibitem{Tycetal:2010}
T. Tyc, H. Chen, C. T. Chan, and U. Leonhardt, IEEE J. of Select. Topics Quantum Electron. {\bf 16}, 418 (2010).


\bibitem{Tycetal:2008}
T. Tyc and U. Leonhardt, New J. Phys. {\bf 10}, 115038 (2008).

\bibitem{Danner:2010}
A. J. Danner, New J. Phys. {\bf 12}, 113008 (2010).


\bibitem{Leonhardt:2011a}
J. Perczel and U. Leonhardt, to be published, preprint available at arXiv (2011).

\bibitem{Milonni:2005}
P. W. Milonni, 
{\it Slow Light, Fast Light and Left-Handed Light}
(Institute of Physics, Bristol and Philadelphia, PA 2005).




\end{thebibliography}
\end{document}